\documentclass[proceedings]{JHEP3}

\PrHEP{PrHEP hep2001}
\conference{International Europhysics Conference on HEP}  

\def\bq{\begin{eqnarray}}
\def\eq{\end{eqnarray}}
\def\l{\langle}
\def\r{\rangle} 
\def\eps{\varepsilon}

\title{
  Steps towards the QCD calculation for $e^+ e^- \rightarrow
  \;\mbox{3 jets}$ at NNLO
  }

\author{Sven Moch, Peter Uwer\\
        Institut f{\"u}r Theoretische Teilchenphysik, Universit{\"a}t Karlsruhe,\\
        76128 Karlsruhe, Germany\\
        E-mail: \email{moch@particle.uni-karlsruhe.de}, \email{uwer@particle.uni-karlsruhe.de}}

\author{\speaker{Stefan Weinzierl}\\
        Dipartimento di Fisica, Universit\`a di Parma, INFN Gruppo Collegato di Parma,\\
        43100 Parma, Italy\\ 
        E-mail: \email{stefanw@fis.unipr.it}}

\abstract{
High precision analyses of experimental data for $e^ +e^-$ annihillation, 
such as determination of jet rates or event shape observables, call for
complete next-to-next-to-leading order (NNLO) perturbative QCD predictions. 
In this talk, we discuss the various ingredients entering the
calculation of the NNLO corrections in $e^ +e^-$ annihillation. 
         }

\begin{document}

\section{Introduction}
Perturbation theory is a powerful tool for precise theoretical predictions
on the outcome of high-energy experiments.
The state-of-the-art is the transition to fully differential 
next-to-next-to-leading order (NNLO) calculations for jet physics.
The most prominent processes where a complete NNLO calculation is desirable are
Bhabha scattering, $pp \rightarrow 2 \;\mbox{jets}$ and
$ e^+ e^- \rightarrow 3 \; \mbox{jets}$.
A NNLO calculation of $e^+ e^- \rightarrow \;\mbox{3 partons}$
is expected to reduce
the theoretical uncertainty in the extraction of $\alpha_s$ down
to 1\% \cite{Burrows:1996db}. Furthermore, it allows to 
model the jet structure more accurately and should improve the knowledge on the
interplay between perturbative and power corrections.

In the past years there has been tremondous progress towards this goal:
The relevant master integrals for $p p \rightarrow 2\;\mbox{jets}$ and
$e^+ e^- \rightarrow 3 \;\mbox{jets}$ have been calculated, either with the help of a
Mellin-Barnes representation \cite{Smirnovetal}
or by solving differential equations \cite{GehrmannRemiddi}.
Numerical results for these master integrals 
serve as a useful cross-check \cite{Binoth:2000ps,Laporta:2000dc}.
With these results the two-loop amplitudes for
Bhabha scattering \cite{Bern:2000ie}, for $p p \rightarrow 2\;\mbox{jets}$ 
\cite{Bern:2000dn,Anastasiouetal}, for $p p \rightarrow \gamma \gamma$ \cite{Bern:2001dg} 
and for light-by-light scattering~\cite{Bern:2001df} have been calculated.

However, the two-loop amplitudes are only one part of the story.
Loop amplitudes in a massless theory like QCD lead to infrared singularities.
These divergences cancel against corresponding singularities resulting from amplitudes
with additional unresolved partons.
Presently, the singular behaviour of double-unresolved tree amplitudes 
\cite{Campbell:1998hg,Catani:1998nv}
and of one-loop amplitudes with one unresolved parton 
is understood \cite{BernDelDucaetal,Kosower:1999rx}, 
as well as the structure of the poles of two-loop amplitudes up to
$1/\eps$ \cite{Catani:1998bh}.
But despite this progress there is still a fair amount of work to be done:
\begin{itemize}

\item The two-loop amplitude for $e^+ e^- \rightarrow 3 \;\mbox{jets}$
remains to be calculated, although recently there has been reported 
progress \cite{NigelandCO}. Due to multiple scales, the amplitude will depend on
the ratio of two kinematical invariants, 
for example $s_{12}/s_{123}$ and $s_{23}/s_{123}$. In contrast, the two-loop
amplitudes calculated so far depend only on one ratio of invariants, say $s/t$.  

\item A method to cancel IR divergences has to be set up. This requires the extension
of the subtraction or slicing method to NNLO. In particular analytic integrations
over unresolved regions in phase space have to be carried out.

\item The final numerical computer program requires stable and efficient
numerical methods, in particular for the Monte Carlo integration of the double unresolved
contributions.
\end{itemize}

In this talk we focus on the first two problems.

\section{Two-loop amplitudes}

Let us briefly review how the calculation of the two-loop amplitudes 
for the single scale problems has been performed.
Starting from the Feynman diagrams, which yield two-loop tensor
integrals, i.e. integrals with powers of the loop momentum in the
numerator, one performs the tensor reduction.
Using a Schwinger parametrization these tensor integrals can be related to scalar integrals
with higher powers of the propagators and different values of the 
dimension $D$ \cite{Tarasov:1996br}.
In a second step, integration-by-parts \cite{Tkachov:1981wb}, 
Lorentz-invariance identities \cite{GehrmannRemiddi}
or form-factor relations \cite{Moch:1999eb}
can be used to eliminate propagators and to reduce the integrals to a few
basic topologies.
At this stage, the basic topologies occur in various dimensions and with various powers $\nu_i$
of the propagators.
The last step relates these basic topologies to the (much smaller) set of master integrals, which
are in most cases just the basic topologies for $D=4-2\eps$ and all $\nu_i=1$.
Again, integral relations derived with integration-by-parts or
form-factor relations are used for this step.
Looking at the computational cost, one finds that the first two steps are rather 
easy to perform, whereas the third step requires to solve large systems of equations.
In particular in the presence of multiple scales this becomes rather involved.

We therefore would like to advocate a different approach, where basic topologies
with arbitrary powers of the propagators and arbitrary dimensions are 
evaluated directly \cite{Moch:2001zr}. 
As an advantage, this approach keeps the size of the intermediate expressions much smaller 
and allows for a very efficient implementation of the reduction scheme. 
The starting point is the observation that these integrals can be written as nested
sums involving Gamma functions. 
For example, a basic integral occuring in $e^+ e^- \rightarrow 3 \;\mbox{jets}$ is the
so-called C-topology, which can be written as \cite{Moch:2001zr}
\bq
\label{result_Ctopo}
I & = & 
 \frac{\Gamma(2m-2\eps-\nu_{1235})\Gamma(1+\nu_{1235}-2m+2\eps)
       \Gamma(2m-2\eps-\nu_{2345})\Gamma(1+\nu_{2345}-2m+2\eps)
      }{\Gamma(\nu_1) \Gamma(\nu_2) \Gamma(\nu_3) \Gamma(\nu_4) \Gamma(\nu_5)
        \Gamma(3m-3\eps-\nu_{12345})} \nonumber \\
& & 
 \times
 \frac{ \Gamma(m-\eps-\nu_5) \Gamma(m-\eps-\nu_{23})}{\Gamma(2m-2\eps-\nu_{235})}
 \left( -s_{123} \right)^{2m-2\eps-\nu_{12345}}
 \sum\limits_{i_1=0}^\infty
 \sum\limits_{i_2=0}^\infty
 \frac{x_1^{i_1}}{i_1!}
 \frac{x_2^{i_2}}{i_2!} \nonumber \\
& & \times \left[
   \frac{\Gamma(i_1+\nu_3) \Gamma(i_2+\nu_2) \Gamma(i_1+i_2-2m+2\eps+\nu_{12345})
                                             \Gamma(i_1+i_2-m+\eps+\nu_{235})
        }{\Gamma(i_1+1-2m+2\eps+\nu_{1235}) \Gamma(i_2+1-2m+2\eps+\nu_{2345})
          \Gamma(i_1+i_2+\nu_{23})}
 \right. \nonumber \\
 & &
 - x_1^{2m-2\eps-\nu_{1235}} 
  \nonumber \\
 & & \times
   \frac{\Gamma(i_1+2m-2\eps-\nu_{125}) \Gamma(i_2+\nu_2) \Gamma(i_1+i_2+\nu_4)
                                             \Gamma(i_1+i_2+m-\eps-\nu_1)
        }{\Gamma(i_1+1+2m-2\eps-\nu_{1235}) \Gamma(i_2+1-2m+2\eps+\nu_{2345})
          \Gamma(i_1+i_2+2m-2\eps-\nu_{15})}
 \nonumber \\
 & &
 - x_2^{2m-2\eps-\nu_{2345}}
 \nonumber \\
 & & \times
   \frac{\Gamma(i_1+\nu_3) \Gamma(i_2+2m-2\eps-\nu_{345}) \Gamma(i_1+i_2+\nu_1)
                                             \Gamma(i_1+i_2+m-\eps-\nu_4)
        }{\Gamma(i_1+1-2m+2\eps+\nu_{1235}) \Gamma(i_2+1+2m-2\eps-\nu_{2345})
          \Gamma(i_1+i_2+2m-2\eps-\nu_{45})}
 \nonumber \\
 & & 
 + x_1^{2m-2\eps-\nu_{1235}} x_2^{2m-2\eps-\nu_{2345}}
   \frac{\Gamma(i_1+2m-2\eps-\nu_{125}) \Gamma(i_2+2m-2\eps-\nu_{345}) 
        }{\Gamma(i_1+1+2m-2\eps-\nu_{1235}) \Gamma(i_2+1+2m-2\eps-\nu_{2345})
          } 
 \nonumber \\
 & & \left. \times
 \frac{
                                          \Gamma(i_1+i_2+2m-2\eps-\nu_{235})
                                          \Gamma(i_1+i_2+3m-3\eps-\nu_{12345})
      }{\Gamma(i_1+i_2+4m-4\eps-\nu_{12345}-\nu_5)}
 \right],
\eq
where we set $x_1=(-s_{12})/(-s_{123})$ and $x_2=(-s_{23})/(-s_{123})$.
Here, the result for the integral holds for arbitrary dimensions 
$D=2m-2\eps$ and any (not necessarily integer) power $\nu_i$ of the propagators.

The Gamma functions can be expanded systematically in $\eps$,
thus allowing to solve these nested sums algorithmically to any given 
order in $\eps$ and to express the result in the basis of $Z$-sums, defined by
\bq 
  Z(n;m_1,...,m_k;x_1,...,x_k) & = & \sum\limits_{n\ge i_1>i_2>\ldots>i_k>0}
     \frac{x_1^{i_1}}{{i_1}^{m_1}}\ldots \frac{x_k^{i_k}}{{i_k}^{m_k}}.
\eq
Important special cases of this definition for $Z$-sums are multiple polylogarithms
(in which we express our final results)
 \cite{Goncharov} - \cite{Remiddi:1999ew} 
\bq
\label{multipolylog}
\mbox{Li}_{m_k,...,m_1}(x_k,...,x_1) & = & Z(\infty;m_1,...,m_k;x_1,...,x_k)  
\eq
and Euler-Zagier sums (which occur in the expansion of Gamma functions)
\bq
Z_{m_1,...,m_k}(n) & = & Z(n;m_1,...,m_k;1,...,1).
\eq
$Z$-sums form an algebra.
In fact, $Z$-sums can be considered as generalizations of Euler-Zagier sums or harmonic
sums.
Many algorithms known for the latter \cite{Vermaseren:1998uu}
carry over to $Z$-sums.
The usefulness of the $Z$-sums lies in the fact, that they interpolate between
multiple polylogarithms and Euler-Zagier sums, 
the interpolation being compatible with the algebra structure.

\section{Subtraction method}

The NNLO cross section receives contributions from double unresolved tree amplitudes, 
one-loop amplitudes with one unresolved parton and two-loop amplitudes:
\bq
d\sigma_{n+2}^{(0)} & = & 
 \left( \left. {\cal A}_{n+2}^{tree} \right.^\ast {\cal A}_{n+2}^{tree} \right) \theta_{n+2}
d\phi_{n+2},  \nonumber \\
d\sigma_{n+1}^{(1)} & = & 
 \left( 
 \left. {\cal A}_{n+1}^{tree} \right.^\ast {\cal A}_{n+1}^{1-loop} 
 + \left. {\cal A}_{n+1}^{1-loop} \right.^\ast {\cal A}_{n+1}^{tree} \right) \theta_{n+1} 
d\phi_{n+1}, \nonumber \\
d\sigma_n^{(2)} & = & 
 \left( 
 \left. {\cal A}_n^{tree} \right.^\ast {\cal A}_n^{2-loop} 
 + \left. {\cal A}_n^{2-loop} \right.^\ast {\cal A}_n^{tree}  
 + \left. {\cal A}_n^{1-loop} \right.^\ast {\cal A}_n^{1-loop} \right) \theta_n d\phi_n,
\eq
where $\theta_n$ is the jet-defining function for $n$ partons.
Taken separately, each part gives a divergent contribution. Only the sum of
all contributions is infrared finite.
This problem already occurs in NLO calculations and can be solved by adding
and subtracting a suitable chosen term. For example
the NLO cross section for $(n+1)$ partons is written as \cite{Catani:1997vz}
\bq
\int d\sigma_{n+2}^{(0)} + \int d\sigma_{n+1}^{(1)} & = &
\int \left( d\sigma_{n+2}^{(0)} - d\sigma^A_{n+1} \right) 
+ \int \left( d\sigma_{n+1}^{(1)} + d\sigma^A_{n+1} \right).
\eq
Here $d\sigma_{n+1}^A$ acts as a local counterterm for single unresolved configurations 
to $d\sigma_{n+2}^{(0)}$ in $D$ dimensions and is
integrable over a one-parton subspace.

A similar subtraction scheme is needed for the NNLO cross section. 
The more complicated part is the subtraction term to $d\sigma_{n+2}^{(0)}$ 
for double unresolved configurations.
In addition, the one-loop corrections $d\sigma_{n+1}^{(1)}$
require a subtraction term $d\sigma_n^{A,loop}$ which approximates 
in $D$ dimensions the one-loop
corrections when one parton becomes unresolved.
We shortly outline how to obtain the subtraction terms for 
$d\sigma_n^{A,loop}$.
We first decompose the amplitudes into primitive amplitudes (e.g. with a fixed
cyclic ordering and a definitive routing of the fermion lines).
Primitive one-loop amplitudes factorize in the collinear limit 
where one parton becomes unresolved as \cite{BernDelDucaetal,Kosower:1999rx}
\bq
\lefteqn{
\left. {\cal A}_{n+1}^{tree} \right.^\ast {\cal A}_{n+1}^{1-loop} 
 + \left. {\cal A}_{n+1}^{1-loop} \right.^\ast {\cal A}_{n+1}^{tree} 
 \rightarrow } \nonumber \\
& & \mbox{Sing}^{tree} \left(
\left. {\cal A}_{n}^{tree} \right.^\ast {\cal A}_{n}^{1-loop} 
 + \left. {\cal A}_{n}^{1-loop} \right.^\ast {\cal A}_{n}^{tree}
\right)
+ \mbox{Sing}^{1-loop} \left. {\cal A}_{n}^{tree} \right.^\ast {\cal A}_{n}^{tree}
\eq
The singular function $\mbox{Sing}^{tree}$ is already known from NLO calculations
and poses no problem.
The function $\mbox{Sing}^{1-loop}$ gives a new contribution. 
However, it is relatively simple to write down an appropriate subtraction term 
for each primitive structure. 
As an example we consider
the splitting $g \rightarrow \bar{q} q$. 
For the primitive part where the fermion entering the loop turns right we have 
as subtraction term 
\bq
\lefteqn{
\l \mu | V^{1-loop}_{\bar{q}_i q_j,k} | \nu \r = } \nonumber \\ 
& & \left[ -g^{\mu\nu} - \frac{4}{2p_ip_j} S^{\mu\nu} \right]
  c_\Gamma 
  \left( \frac{\mu^2}{-2 p_i p_j} \right)^\eps 
  \left\{ 
           \frac{1}{\eps^2} + \frac{3}{2} \frac{1}{\eps(1-2\eps)}
                          + \frac{1}{1-2\eps}
  \right\} 
\eq
Here $S^{\mu\nu}$ denotes the spin correlation tensor.
This subtraction term has to integrated over a one-parton phase space.
For the more complicated integrals we can rely on the technique of nested sums \cite{Phaf:2001gc}.
The integrated counter-part reads:
\bq
\lefteqn{
\int d\phi_{dipole} \frac{1}{2p_ip_j} V^{1-loop}_{\bar{q}_i q_j,k} 
 =  \left( - g^{\mu\nu} \right) 
 c_\Gamma^2 \left( \frac{\mu^2}{P^2} \right)^\eps \left( \frac{\mu^2}{-P^2} \right)^\eps 
} \nonumber \\
& & \cdot 
  \left\{ 
   - \frac{1}{3} \frac{1}{\eps^3}
   - \left( \frac{23}{18} + \frac{1}{3}\gamma \right) \frac{1}{\eps^2}
   - \left( \frac{1}{6} \gamma^2 + \frac{23}{18} \gamma + \frac{118}{27} \right) \frac{1}{\eps}
\right. \nonumber \\ & & \left.
   - \left( \frac{7}{9} \zeta_3 + \frac{1}{18} \gamma^3 + \frac{23}{36} \gamma^2
            + \frac{118}{27} \gamma + \frac{1}{6} \pi^2 + \frac{4075}{324} \right)
  \right\}
   + O(\eps)
\eq
We note that in general for one-loop amplitudes with one unresolved parton the soft
radiation pattern is more complicated as compared to a NLO
calculation. Generally, a simple dipole picture with emitter and
spectator will not be sufficient.

\section{Outlook and conclusions}

The NNLO calculation for $ e^+ e^- \rightarrow 3 \; \mbox{jets}$ is a 
challenging project.
We have reported on recent progress in the evaluation of two-loop 
amplitudes with multiple scales using the novel approach of nested sums.
All algorithms for the solution of the nested sums are suitable for
implementation in computer algebra systems 
like GiNaC \cite{Bauer:2000cp} or FORM \cite{Vermaseren:2000nd}.
We have outlined the road to the cancellation of infrared divergences.
This involves the analytic integration over the single and double unresolved 
partonic phase space and we believe the technique of nested sums to be 
of great value here as well. 
We have not addressed any issues \cite{Weinzierl:1999yf} concerning the final numerical
integration of the fully differential NNLO cross section 
by means of Monte Carlo techniques.

\end{document}